\documentclass[pra,aps,floatfix,showpacs]{revtex4-1}
\usepackage{epsfig,graphicx,tabularx,epstopdf}
\newenvironment{DIFnomarkup}{}{}
\begin{document}
\title{Multimode model for an atomic Bose-Einstein condensate in a ring-shaped optical lattice}
\author{D. M. Jezek and H. M. Cataldo}
\affiliation{IFIBA-CONICET
Pabell\'on 1, Ciudad Universitaria, 1428 Buenos Aires, Argentina}
%
%
\begin{abstract}
We study
the population dynamics of a ring-shaped optical lattice with a high
number of particles per site and a low, below ten, number of wells.
Using a localized on-site   basis  defined in terms of stationary states, we 
were able to construct a multiple-mode model depending on relevant hopping and on-site energy 
parameters. We show  that in case of two wells, our model corresponds exactly to  the latest improvement of the two-mode model.
 We derive  a  formula  for the  self-trapping period, which turns out to be
 chiefly ruled by the on-site interaction energy parameter.
By comparing to time dependent Gross-Pitaevskii simulations,
we show that the multimode model results
can be enhanced  in a remarkable way
   over all the regimes by only renormalizing such a  parameter. 
Finally, using a different approach which involves only the ground
state density, we derive an effective interaction energy parameter that shows
to be in accordance with the renormalized one.

\end{abstract}
\pacs{03.75.Lm, 03.75.Hh, 03.75.Kk}

\maketitle
\section{Introduction}

The two-mode model applied to double-well atomic Bose-Einstein condensates
 has been extensively studied in the last years 
\cite{smerzi97,ragh99,anan06,jia08,albiez05,mele11,abad11,doublewell,*xiong,*zhou,*cui,*gui,
jez13,*cond}
Assuming that the order parameter can 
be described as a 
superposition of localized on-site wave functions  with time dependent coefficients,
such a model predicts Josephson and self-trapping regimes \cite{smerzi97,ragh99}, which have 
been  experimentally
observed by Albiez {\it et al.}
\cite{albiez05}. 

The self-trapping (ST) phenomenon, which
is also present in extended optical lattices \cite{optlat,*xue,*alex,*fu}, is a non linear effect where an
 initially highly populated (over a critical value) site,
remains with a larger number of particles than the remaining sites
 over all the evolution. 
There is nowadays an active research on the self-trapping effect, 
which involves different types of systems, including mixtures of atomic species \cite{stlastoplat,*Adhi}.

The dynamics of ring-shaped optical lattices 
with three \cite{trespozos2011} and four wells
 \cite{cuatropozos06},  has been previously
investigated through multiple-mode (MM) models which utilized {\em ad-hoc} 
values for hopping and on-site energy parameters.
In the present article instead, we will extract such parameters
 from a mean-field approach using localized on-site  functions.
We  have shown
in a previous work \cite{cat11}
that in a ring-shaped optical lattice, localized on-site (which  we  called `Wannier-like' (WL)) 
functions can be obtained in terms of
stationary states of the Gross-Pitaevskii (GP) equation  with different winding numbers. 
Here we will show  that the above parameters yield the same type of
corrections to the MM model for large filling factors, 
as those obtained for the improved two-mode (TM) model for two-well systems \cite{anan06}.
 
We will derive an approximate formula for the self-trapping period
in terms of the on-site interaction energy parameter. Using this formula and a single
GP simulation results, 
a renormalizing on-site energy parameter
that substantially improves the MM model  can be obtained, 
in what will be called the renormalized multiple-mode (RMM) model.
Taking into account the density deformation during the time evolution \cite{smerzi03},
it has been shown in a recent work that for a double-well system
an effective interaction energy parameter  should be considered 
in the TM model to properly describe the exact dynamics  
\cite{jez13,*cond}. 
Here we will adapt the same approach to our multiple-well system, which
will allow us to obtain such an
effective parameter only in terms of the ground state density. 
Finally, we will show that both approaches give similar results.

This paper is organized as follows. In Sec. \ref{descrip} we describe the system and 
in Sec. III we
outline the method for obtaining the WL functions, along with the properties required for
building a reasonable  multimode model dynamics. There we also  
define the  model parameters in terms of the WL functions.
 In Sec. IV we specialize to the case of two wells, showing that our treatment
through the WL functions turns out to be
 exactly the same as the latest two-mode model formulation 
\cite{anan06}. Next, by means of the formula derived for the ST period, we show that
the two-mode model can be enhanced in a remarkable way by only renormalizing the on-site interaction energy
parameter. In Sec. V we develop the multiple-mode model, which generalizes our finding of the 
previous section. Finally, based on the method described in Ref. 
\cite{jez13,*cond}, in
Sec. VI we derive an
 effective interaction energy parameter and compare it with the renormalized one.
To conclude, a summary of our work is presented in Sec. VII. 

\section{Ring-shaped lattice and condensate parameters}\label{descrip}
 
We consider  a Bose-Einstein condensate  of Rubidium atoms confined by 
an external trap $ V_{\text{trap}}$, consisting of a superposition of
a toroidal term $V_{\text{toro}}$  and a lattice potential  $V_{\text{L}}(x,y)$
 formed by radial barriers.
Similarly to the trap 
utilized in  recent experiments \cite{ryu07,wei08}, the toroidal trapping  potential
 in cylindrical coordinates reads,
\begin{equation}
V_{\text{toro}}(r,z ) = \frac{ m }{2 } \left[\omega_{r}^2  r^2 
+ \omega_{z}^2  z^2\right] +  
V_0 \, \exp ( -2 \, r^2/\;\lambda_0^2)
\end{equation}
where $m$ is the atom mass and $\omega_{r}$  and $\omega_{z}$ denote the radial and axial 
frequencies, respectively. 
We have set $\omega_z >>
\omega_r$ to suppress excitation in the $z$ direction.
In particular, we have chosen
 $ \omega_r / (2 \pi) =  7.8 $ Hz  and $ \omega_z / (2
\pi) = 173 $ Hz, while for the laser beam we have set 
$ V_0  =  100  \, \hbar \omega_r$  and $ \lambda_0 = 6 \, l_r $, with 
$ l_r = \sqrt{\hbar /( m \omega_r)}$. 
On the other hand, the lattice  potential is formed by $N_c$
Gaussian  barriers  located at equally spaced angular positions 
$ \theta_k = 2 \pi k /N_c $,
where $-[[(N_c-1)/2]]\leq k\leq [[N_c/2]]$ with $[[.]]$ denoting the integer part, 
\begin{equation}
V_{\text{L}}(x,y) =
V_b \,\, \sum_{k=-[[(N_c-1)/2]]}^{[[N_c/2]]} \exp \left\{ - \frac{ [ \cos(\theta_k) \, 
y  - \sin(\theta_k) \,x]^2}  
{ \lambda_b^2}\right\} \Theta[ \sin(\theta_k) y + \cos(\theta_k) x ],
\end{equation}
where $ \Theta $ denotes the Heaviside function. 
For the  numerical calculations  we have fixed the width of the Gaussians  to  $ \lambda_b = 0.5\, l_r $  and the barrier height to $ V_b= 80 \,
\hbar \omega_r $.
In the mean-field approximation, the stationary states are solutions of the 
GP equation \cite{gros61,*pita}
\begin{equation}
\left[-\frac{ \hbar^2 }{2 m}{\bf \nabla}^2  +
V_{\rm{trap}}({\bf r})+g\,N|\psi_n({\bf r})|^2\right]\psi_n({\bf r})
= \mu\,\psi_n({\bf r}),
\label{gp}
\end{equation}
where $ \psi_n({\bf r})$ denotes a 
 two-dimensional (2D) 
order parameter \cite{castin} 
normalized to unity
with winding number $n$ \cite{je11}.
The vorticity is numerically imprinted following the procedure described in Ref.
\cite{nuestro08}.
 $N$ and $\mu$  denote, 
 respectively, the number 
of particles and  the chemical potential ($ N=10^5 $ will be assumed over all the numerical calculations).
The effective 2D coupling constant $g=g_{3D}\sqrt{m\omega_z/2\pi\hbar} $
  is written in terms of the 3D coupling constant between the atoms 
$g_{3D}=4\pi a\hbar^2/m$, where $a= 98.98\, a_0 $ denotes the  
$s$-wave scattering length of $^{87}$Rb, $a_0 $ being the Bohr radius.
Technical advances have been recently achieved, to obtain  experimentally this type of condensates
  in ring-shaped optical lattices with an arbitrary number of sites  \cite{hen09}.

\section{ Localized states and  hopping and on-site energy parameters}

In this section we will summarize the results obtained in a previous work \cite{cat11} that 
will be used to describe the present dynamics.
We are interested in studying the Josephson and ST regimes.
Such a dynamics takes place when
the ground-state chemical potential becomes smaller than
the minimum of the effective potential barrier dividing two lattice sites
\cite{je11}.

\subsection{ Localized WL states}
\begin{figure}
\includegraphics{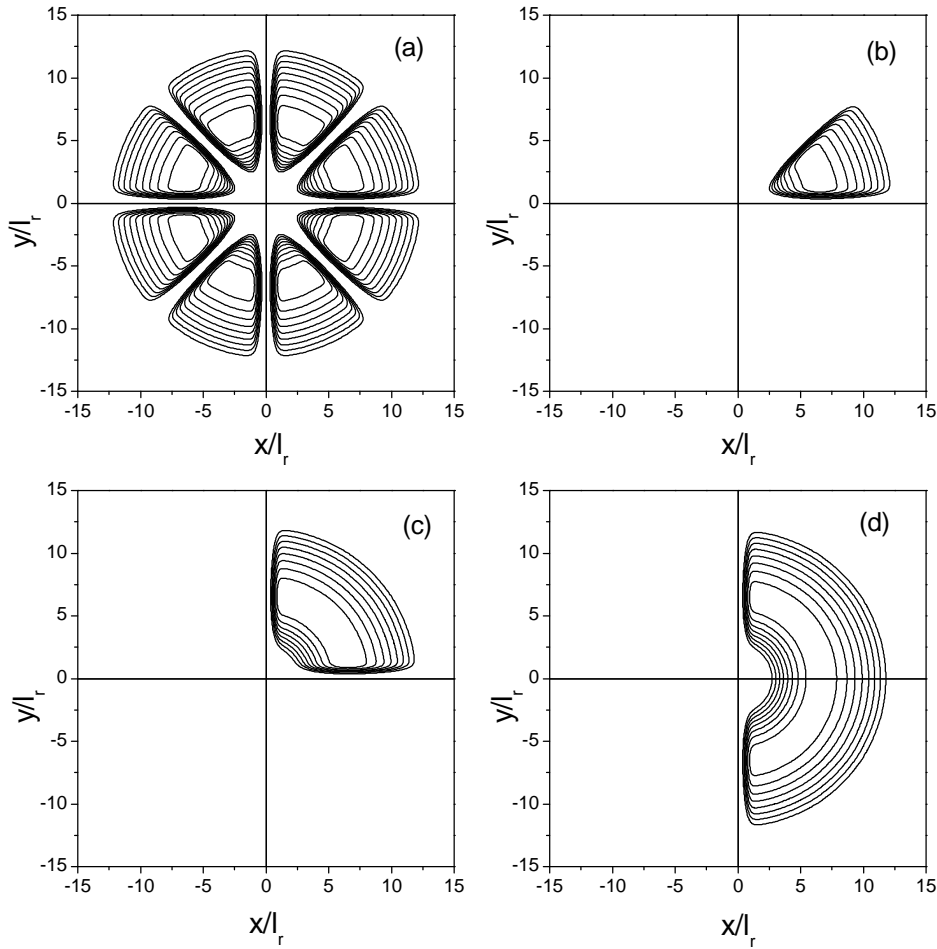}
\caption{Density isocontours of the ground-state wave function  $|\psi_0|^2$ for $N_c=8$ (a),
and of the WL function $ w_0^2({\mathbf r}) $  for the following numbers of sites 
$N_c=8$ (b), 
$N_c=4 $ (c), and $N_c=2$ (d).
}
\label{wan1}
\end{figure}

The stationary states $\psi_n( r, \theta )$ are obtained as the numerical 
solutions of Eq. (\ref{gp}) \cite{je11}.
Assuming   large
barrier heights \cite{cat11},
the winding number $n$ will be restricted to the values 
$-[[(N_c-1)/2]]\leq n \leq [[N_c/2]]$ \cite{je11}. 
 We have seen in Ref. \cite{cat11} that stationary states
of different winding number must be orthogonal,
 and that the following definition for the WL functions 
\begin{equation}
w_k({ r, \theta })=\frac{1}{\sqrt{N_c}} \sum_{n} \psi_n({ r, \theta})
 \, e^{-i n\theta_k } \,,
\label{wannier}
\end{equation}
corresponds indeed to well localized functions on each $k$-site.
In addition, it has been shown in Ref. \cite{cat11} that the above orthogonality implies that
 the set of $N_c$   WL  functions (\ref{wannier}) located at different
$k$-sites, must also form 
an orthonormal set. In Fig.~\ref{wan1} we have depicted the WL function density $ w_0^2$ 
for several values of $N_c$, 
where it becomes clear
that they  are certainly well-localized functions. 
Here it is important to recall that the main difference between our WL function and a
`true' Wannier function consists in that only the former depends on the
filling factor, i.e. the number of particles at each site, as seen
in Ref. \cite{cat11}. 
One may also write  the above stationary states in terms of these 
localized functions,
\begin{equation}
\psi_n({\mathbf r}) =\frac{1}{\sqrt{N_c}} \sum_{k} w_k ({ r, \theta})
 \, e^{i n \theta_k } \,,
\label{inwannier}
\end{equation}
which  will be useful to describe the dynamics we are interested in. 

\subsection{ Hopping and on-site energy parameters}\label{parameters} 

The $N_c$-mode dynamics will be described in terms of  the following parameters,
\begin{equation}
\varepsilon = \int d^2{\bf r}\,\, w_0( r, \theta)\left[
-\frac{ \hbar^2 }{2 m}{\bf \nabla}^2 +
V_{\rm{trap}}({\bf r})\right]w_0( r, \theta)
\label{eps0}
\end{equation}
\begin{equation}
J= -\int d^2{\bf r}\,\, w_0( r, \theta)\left[
-\frac{ \hbar^2 }{2 m}{\bf \nabla}^2  +
V_{\rm{trap}}({\bf r})\right]w_1( r, \theta)
\label{jota0}
\end{equation}
\begin{equation}
J'= -2\,g\int d^2{\bf r}\,\, w_0^3( r, \theta)
\,w_1( r, \theta)
\label{jotap0}
\end{equation}
\begin{equation}
U= g \int d^2{\bf r}\,\, w_0^4( r, \theta),
\label{U0}
\end{equation}
which due to the symmetry of the lattice can be written without loss of generality only in terms of
 the $k=0$ and $k=1$ sites. 
We want to mention that these parameters can also be efficiently evaluated through
the alternative formulae given in Ref. \cite{cat11}. 

\section{Two-mode dynamical equations}\label{2-mode}

When the trapping potential consists of a double well, the condensate dynamics
may be simply described through a pair of coupled equations, which corresponds to the two-mode model.
Such a TM dynamics has been extensively studied in recent years \cite{ragh99}. 
Particularly, an improved version of this model \cite{anan06} has been also applied
to particles exhibiting a
 dipolar interaction, which generates self-induced Josephson junctions in a similar toroidal 
geometry \cite{abad11}.

\subsection{Dynamical equations in terms of the coefficients of well-localized functions}\label{2-model}

The commonly used ansatz for the TM  wave function reads 
\begin{equation}
\psi_{\rm TM}(r,\theta,t)= b_R(t)\, \psi_R(r,\theta) + b_L(t)\, \psi_L(r,\theta),
\label{varan2m}
\end{equation}
where $ \psi_R(r,\theta)$ and $ \psi_L(r,\theta)$  are well-localized functions at 
the right and left well, respectively.
Such wave functions are easily identified with the WL functions, namely
$ \psi_R(r,\theta)= w_0 (r,\theta )$ and $ \psi_L(r,\theta)= w_1 (r,\theta )$, 
since from Eq. (\ref{wannier}) 
we get for $N_c=2$,
\begin{equation}
w_0({ r, \theta })=\frac{1}{\sqrt{2}} ( \psi_0({ r, \theta}) + \psi_1({ r, \theta})) \,,
\label{2wannier0}
\end{equation}
\begin{equation}
w_1({ r, \theta })=\frac{1}{\sqrt{2}} ( \psi_0({ r, \theta}) - \psi_1({ r, \theta})) \,,
\label{2wannier1}
\end{equation}
which turns out to be identical to the standard  TM variational proposal \cite{smerzi97}. 

Note that the first excited 
state is an odd function of $x$,  
as required by the TM  model (antisymmetric solution). 
This can be easily verified by noting that the stationary state with winding number
$n=1$ 
has uniform phases at the right and left 
well with values $\phi=0$ and $\phi= \pi$, respectively \cite{je11}. 
Here we may recall that this 
 only occurs in the regime of large barriers \cite{cat11}, 
where $ \psi_1( {\mathbf r} )= \psi_{-1}( {\mathbf r} )$ may be taken as a real function 
that does not carry any angular momentum,
and hence does not correspond to a `vortex' state \cite{je11}.

In order to obtain the TM equations, we may replace
 the following order parameter, which is written in  terms of the  WL 
functions according to the ansatz (\ref{varan2m}),
\begin{equation}
\psi_{\rm TM}(r,\theta,t)= b_0(t) \, w_0( r, \theta ) + b_1(t) \, w_1( r, \theta ) \, ,
\label{varan2mw}
\end{equation}
in the time dependent GP equation,
\begin{equation}
i\hbar\frac{\partial\psi_{\rm TM}(r,\theta,t)}{\partial t}=
\left[-\frac{ \hbar^2 }{2 m}{\bf \nabla}^2  +
V_{\rm{trap}}(r,\theta)+g\,N|\psi_{\rm TM}(r,\theta,t)|^2\right]\psi_{\rm TM}(r,\theta,t) \, .
\label{2t-dgp}
\end{equation}
Making use of the orthonormality of the WL  functions and recalling the definitions of
hopping and on-site energy parameters (Eqs. (\ref{eps0}) to
 (\ref{U0})) one obtains,
\begin{equation}
i\hbar\,\frac{db_0}{dt} =  \varepsilon b_0-Jb_1+UN|b_0|^2b_0
-\frac{J'}{2}N\,[2 {\rm Re}(b_0^*b_1)b_0
+ b_1],
\label{2mode0}
\end{equation}
\begin{equation}
i\hbar\,\frac{db_1}{dt} =  \varepsilon b_1-Jb_0+UN|b_1|^2b_1
-\frac{J'}{2}N\,[2 {\rm Re}(b_1^*b_0)b_1
+ b_0].
\label{2mode1}
\end{equation}
The above
equations  correspond to the improved TM model developed in Ref. \cite{anan06} and
 applied in Refs. \cite{mele11,abad11}. Here it is worth noticing that
we have disregarded in our derivation terms of the order of the following integral
\begin{equation}
I=g N \int d^2{\bf r}\,\, w_0^2( r, \theta)
\,w_1^2( r, \theta),
\label{iw2}
\end{equation}
since the corresponding contributions have been shown to be negligible,
as also been argued in Ref. \cite{anan06} for high barriers. 

\subsection{Dynamical equations in terms of the particle imbalance and phase difference}\label{2-modelimba}

A more convenient set of variables is obtained by observing  
that $ b_k(t)=  |b_k(t)|  e^{i \phi_k(t)} $, where $ \phi_k(t) $ is the uniform phase of the $k$-site 
and  $ n_k= N_k(t)/ N = |b_k(t)|^2 $ denotes the corresponding filling factor. 
 Following the same procedure of Ref. \cite{abad11}, the equations of motion for the
conjugate coordinates, namely imbalance $ Z = n_0 - n_1 $ and phase difference
 $ \varphi= \phi_1- \phi_0$ read,
\begin{equation}
 \frac{dZ}{dt} = - \sqrt{1-Z^2}\,\sin\varphi 
\label{imb}
\end{equation}
\begin{equation}
 \frac{d\varphi}{dt} =  \Lambda_{\rm eff} Z + \left[
\frac{Z}{\sqrt{1-Z^2}}\right]\cos\varphi,
\label{phase}
\end{equation}
where the time $t$  (in the derivatives)  has been expressed in units of $\hbar/2J_{\rm eff}$
and we have defined $ \Lambda_{\rm eff} = \frac{U N}{2J_{\rm eff}} $, being $ J_{\rm eff}  = J + \frac{ J' }{2}\,N$.
Note that the above equations possess the same structure as the standard TM ones, except that
the bare $J$ has been replaced by an effective hopping parameter $ J_{\rm eff}$, which takes into account the
interaction between particles. 
It is interesting to 
recall that $J$
 may be negative, as occurs in the present calculations,  while $ J_{\rm eff}$ remains always positive \cite{cat11}.

The TM equations (\ref{imb}) and (\ref{phase}) can also be derived from
the following `classical' Hamiltonian,
\begin{equation}
 H(Z,\varphi) = \frac{1}{2}\Lambda_{\rm eff} Z^2 - \sqrt{1-Z^2}\cos\varphi  \ ,
\label{Ham}
\end{equation}
since we have
\begin{equation}
 \frac{dZ}{dt}=-\frac{\partial H}{\partial\varphi} \qquad\qquad \frac{d\varphi}{dt}=
\frac{\partial H}{\partial Z} \ .
\end{equation}
For low $  \Lambda_{\rm eff} $ values the Hamiltonian exhibits only a minimum at $(Z,\varphi)=(0,0)$ and the 
dynamics becomes restricted to Josephson type oscillations.
 For $  \Lambda_{\rm eff} > 1 $ a maximum appears
at $\varphi= \pi $ and 

\begin{equation}
 Z_M = \sqrt{ 1  - \frac{1}{\Lambda_{\rm eff}^2 }} \, ,
\end{equation}
which gives rise to a self-trapping regime. Around this maximum the orbits are
restricted to only one sign of the imbalance. In other words, if one starts with a
positive imbalance it always remains positive.
A ST running-phase mode \cite{smerzi97} arises for $ \Lambda_{\rm eff} > 2 $, which is characterized
by an unbounded $ \varphi $ value. To  find the value $Z_c$ above which
the dynamics becomes ST for
 $\varphi(t=0)=0$, we need to impose the condition $H(Z_c,0)=H(0,\pi)$, which yields
\begin{equation}
 Z_c = 
2\frac{\sqrt{\Lambda_{\rm eff}-1 }}{\Lambda_{\rm eff} } \ .
\label{z_c}
\end{equation}

In this work we are interested in the range $ \Lambda_{\rm eff} >> 1 $ and thus a small $Z_c$ value is attained.
In fact, we have calculated the values of on-site energy and hopping parameters,
$ U = 6.73 \times 10^{-4} \, \hbar \omega_r $, 
$ J = -3.66 \times 10^{-4} \hbar \omega_r $, and
$ \frac{ J' }{2}\,N  = 5.05  \times 10^{-4} \, \hbar \omega_r $, 
from which we obtained  the TM parameters,
$ J_{\rm eff} = 1.39 \times 10^{-4} \, \hbar \omega_r $ and
$ \Lambda_{\rm eff} = 2.42 \times 10^{5}  $.  
In addition, we have found 
 $I=6.05  \times 10^{-7} \, \hbar \omega_r $ (Eq. (\ref{iw2})), which justifies having neglected
terms proportional to such a parameter in the equations.

\begin{figure}
\includegraphics{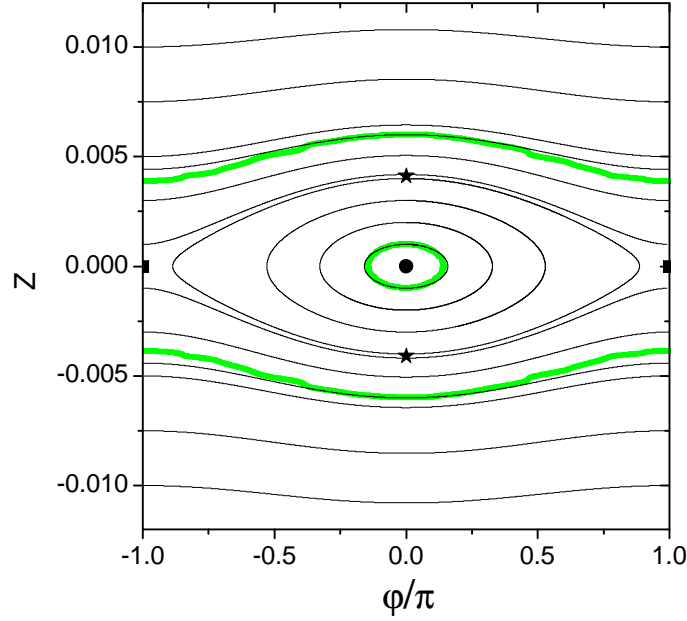}
\caption{(Color online) Phase diagram of the improved TM model for imbalance 
$ Z $  and phase difference  $ \varphi $. 
 The circle and square points respectively indicate the positions of 
the minimum and saddle points of the Hamiltonian (\ref{Ham}),
while the star points correspond to the critical  value $|Z_c|$ (\ref{z_c}). 
We have also depicted as thicker (green) solid lines 
the results of the GP simulation for 
the initial conditions $ (|Z|, \varphi)=(0.001,0) $  and  $ (|Z|, \varphi)=(0.006, 0 )$.} 
\label{fig2}
\end{figure}

In Fig. \ref{fig2} we show the phase diagram $ (Z, \varphi) $ for $ |Z| < 0.012 $, since for 
larger values of $ |Z| $ the
orbits are almost horizontal. The thicker (green) lines correspond  to exact numerical evolutions for 
the initial
conditions: $ (|Z|, \varphi)=(0.001,0) $ (Josephson) and  $ (|Z|, \varphi)=(0.006, 0 )$ (ST).
To obtain such simulations, we have solved the time dependent GP equation with an initial wave function
which reproduces the same initial condition assumed for the TM model evolution. 
In order to compare the phase differences of both
results, we have averaged the GP phase in each $k$-well according to,
\begin{equation}
\phi_k=  \int d^2{\bf r}\,
\, w_k^2({\mathbf r} ) \phi( {\mathbf r}) \, ,
\label{meanphase}
\end{equation}
where $ \phi( {\mathbf r})$ denotes the phase of the GP wave function. 
We want to note that the Bloch states of Ref. \cite{je11} correspond to
equally populated wells with different winding numbers.
 In the double-well potential 
these states are represented by the stationary points  located at $Z=0$ in Fig. \ref{fig2}.
  The minimum at $\varphi=0$ corresponds to a vanishing winding number,
while the saddle point at $ |\varphi|= \pi $ corresponds to winding numbers with $ |n|= 1$.
The latter, however, does not correspond to a vortex state,
since it possesses  zero angular momentum, as discussed in Ref. \cite{je11}.

\begin{figure}
\includegraphics{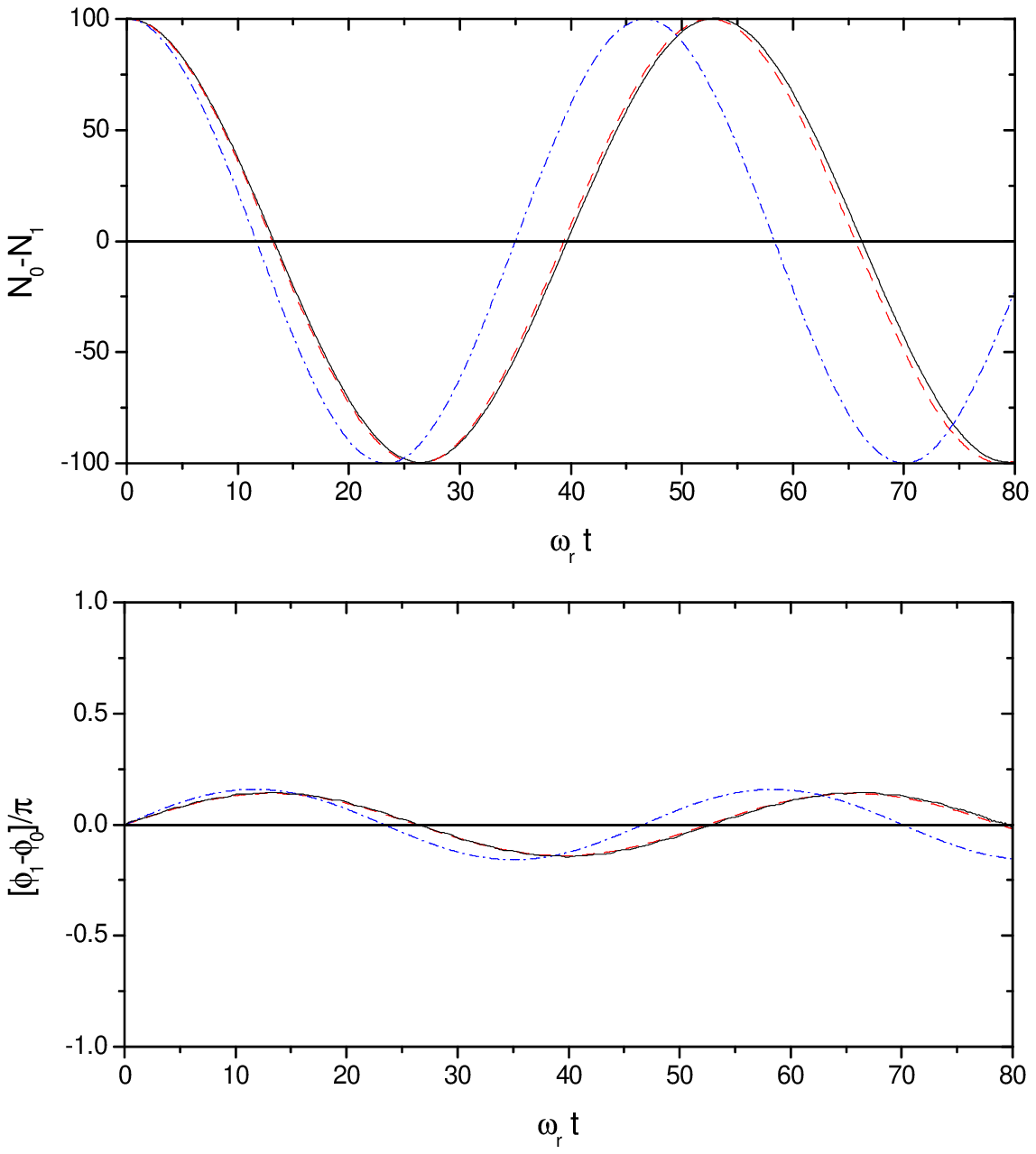}
\caption{(Color online) Josephson oscillation in the double-well system 
with an initial imbalance $Z=10^{-3}$. 
Imbalance (top panel) and phase difference (bottom panel) are depicted as  functions of time. 
The solid line 
corresponds to the GP simulation, while
the dot-dashed (blue) and dashed (red) lines 
correspond to TM evolutions  with $  U = 6.73 \times 10^{-4} \, \hbar \omega_r $  and 
a renormalized on-site energy parameter 
$U_R= 5.28 \times 10^{-4} \, \hbar \omega_r $, respectively.
 }
\label{fig3}
\end{figure}

\begin{figure}
\includegraphics{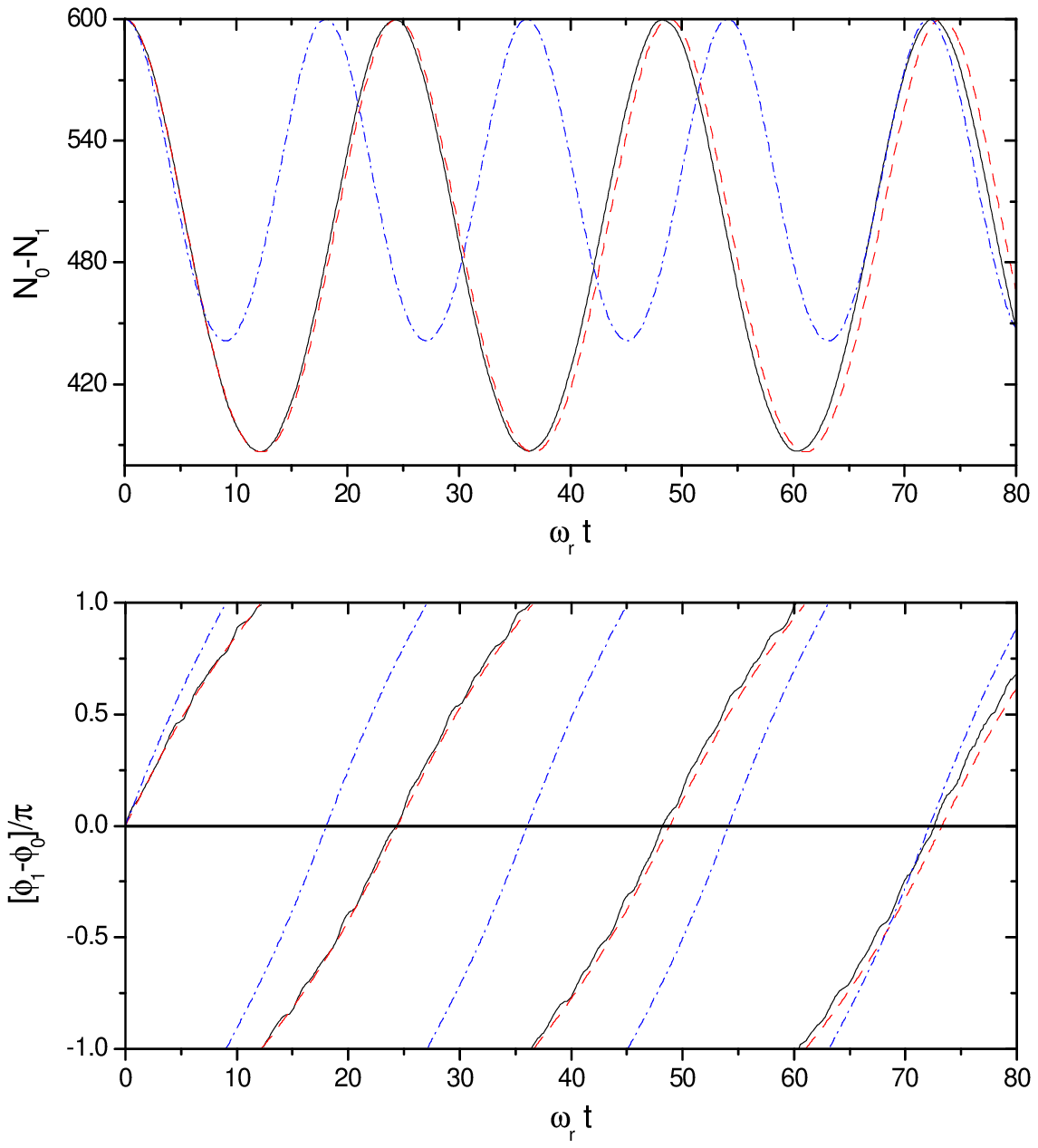}
\caption{(Color online) Same as Fig. \ref{fig3} for a self-trapping evolution with an initial imbalance $Z=6 \times 10^{-3}$.}
\label{fig4}
\end{figure}

Typical time evolutions of Josephson oscillations and ST orbits are shown 
in Figs. \ref{fig3} and \ref{fig4}, respectively. In Fig. \ref{fig3} we depict $Z$ and $\varphi$ as functions of 
time for the GP simulations (solid line), together with the results arising from
the TM model (dot-dashed (blue) line).  On the other hand, Fig. \ref{fig4} shows
the same evolutions for a larger initial imbalance, where we clearly 
observe a self-trapping behavior. In both cases we may see that the TM model predicts a faster
dynamics than the GP simulation. As we will show in the next subsection, 
such a discrepancy can be
substantially reduced when using a renormalized on-site interaction energy parameter (dashed (red) line).

\subsection{ Characteristic times}\label{2-modeltimes}

In this subsection we will derive a formula for the  period of ST oscillations.
Before this, we recall that the Josephson period in the limit
of small oscillations  reads \cite{smerzi97,mele11},
\begin{equation}
T_{\rm so}= \frac{\pi\hbar}{ J_{\rm eff} \sqrt{\Lambda_{\rm eff} + 1}}  \,.
\label{tpeqosc}
\end{equation}
Replacing in  the above equation 
 the values of Sec. \ref{2-modelimba},
we obtain $ T_{\rm so}= 46.0  \,\,\, \omega_r^{-1}$.
This is a rather good estimate of the TM period in Fig. \ref{fig3} ( $ T_{\rm TM}= 46.68 \,\,\, \omega_r^{-1}$),
but it clearly underestimates
the corresponding GP period  ($ T_{\rm GP} \simeq 53,09 \,\,\, \omega_r^{-1}$).

On the other hand, 
the phase difference increases almost
linearly with time  for small imbalance oscillations in the ST regime, 
\begin{equation}
\varphi(t) \simeq \frac{2 \pi}{T_{\rm ST}} t  \,,
\label{fist}
\end{equation}
as seen in Fig. \ref{fig4}.
Now, to be consistent with this approximation, we first rewrite 
  Eq. (\ref{phase}) without the adimensionalized time, and next approximate such an expression
 for $ \Lambda_{\rm eff} = \frac{U N}{2J_{\rm eff}}  >> 1$ and $|Z|<<1$ as follows,
\begin{equation}
 \frac{d\varphi}{dt} =  \frac{N U}{\hbar} Z + \frac{2J_{\rm eff}}{\hbar}\left[
\frac{Z}{\sqrt{1-Z^2}}\right]\cos\varphi \simeq \frac{N U}{\hbar} Z \simeq \frac{N U}{\hbar} Z_0,
\label{Iphaseap}
\end{equation}
where $ Z_0 = \overline{Z(t)} $ denotes the mean value of the time dependent imbalance.
Note in Fig. \ref{fig4} that the maximum departure of $Z(t)$ from such a mean value lies within a $20$ percent.
Therefore, from (\ref{fist}) and (\ref{Iphaseap}) we may estimate the ST period as, 
\begin{equation}
T_{\rm ST}= \frac{ 2 \pi\hbar}{ U \overline{\Delta N }}  \,,
\label{tst}
\end{equation}
where $ \overline{\Delta N} =   Z_0 \,  N $ denotes the time
average of the particle number difference between sites.
If we calculate such an average from the TM model of Fig.~\ref{fig4}, we obtain
$ \overline{\Delta N} \simeq 520 $, from which
Eq. (\ref{tst}) yields
$ T_{\rm ST}= 17.94 \,\, \omega_r^{-1} $,
that is
a good estimate of the TM period of
$18.04 \,\, \omega_r^{-1}$ in Fig. \ref{fig4}. Now, given
that the TM dynamics turns out to be noticeably
faster than the GP evolution, while conserving the shape, it suggests that a 
renormalized value of $U$ in (\ref{tst}) could heal this mismatch. 
In fact, being $ T_{\rm GP}= 24.1 \,\, \omega_r^{-1}$ and 
$ \overline{\Delta N}_{\rm GP} = 494 $, we may propose to replace $U$ in Eq. (\ref{tst}) by
the following renormalized on-site interaction energy parameter:
\begin{equation}
U_{R}= \frac{ 2 \pi\hbar}{  T_{\rm GP}  \overline{\Delta N}_{\rm GP}}
= 5.28 \times 10^{-4} \, \hbar \omega_r \,.
\label{urenor}
\end{equation}
Thus, we have repeated the numerical calculations of the TM model with the above parameter, finding
an excellent agreement with the GP results, as clearly observed in Figs. \ref{fig3} and \ref{fig4}.
It is also remarkable that the period for small Josephson oscillations (\ref{tpeqosc}),
gets now closer to the GP value when using the renormalized parameter (\ref{urenor})
($ T_{\rm so}=51.5 \, \omega_r^{-1}$).  
Therefore, a more accurate Hamiltonian (\ref{Ham}) can be constructed by replacing $ U$ by $U_R$ in
$ \Lambda_{\rm eff}  $.

\section{Multiple-mode dynamical equations}\label{nc-mode}

The  two-mode equations 
describing the boson Josephson junction dynamics of two
weakly coupled Bose-Einstein condensates \cite{ragh99}, along with their recent improvements
for high particle numbers \cite{anan06,jia08},
can be generalized to  multiple-mode (MM) dynamical equations for $N_c$ Bose-Einstein 
condensates forming
a ring.  
In fact, we look for a solution of the time-dependent GP equation
\begin{equation}
i\hbar\frac{\partial\psi_{\rm MM}(r,\theta,t)}{\partial t}=
\left[-\frac{ \hbar^2 }{2 m}{\bf \nabla}^2  +
V_{\rm{trap}}(r,\theta)+g\,N|\psi_{\rm MM}(r,\theta,t)|^2\right]\psi_{\rm MM}(r,\theta,t)
\label{t-dgp}
\end{equation}
within the variational ansatz
\begin{equation}
\psi_{\rm MM}(r,\theta,t)= \sum_{k=0}^{N_c -1}  b_k(t)\,\,w_k(r,\theta),
\label{varan}
\end{equation}
where the phase of the time-dependent complex amplitude $b_k$ corresponds to the uniform phase of the 
order parameter at the $k$-th site, while $N|b_k|^2$ yields the site population.
Then, replacing (\ref{varan}) in (\ref{t-dgp}) and making use of the orthonormality of the
set of  WL  functions, we may extract
 the following system of $N_c$ nonlinear equations,
\begin{eqnarray}
i\hbar\,\frac{db_k}{dt} & = & \varepsilon b_k-J(b_{k-1}+b_{k+1})+UN|b_k|^2b_k
-\frac{J'}{2}N\{2 {\rm Re}[b_k^*(b_{k-1}+b_{k+1})]b_k \nonumber\\
&+& (|b_k|^2+|b_{k-1}|^2)b_{k-1}
+(|b_k|^2+|b_{k+1}|^2)b_{k+1}\}.
\label{ncmode}
\end{eqnarray}
Note that the above expression assumes that each site $k$ is surrounded by two different neighbors
$k-1$ and $k+1$, for that reason  the  $N_c=2$ case has been treated   separately.
In addition, the site denoted by $k=N_c$ ($k=-1$) must be
identified with that of $k=0$
($k=N_c-1$). If we
use $ b_k = |b_k|e^{i \phi_k }$, the time derivative in (\ref{ncmode}) reads
\begin{equation}
i\hbar\,\frac{db_k}{dt} = 
 \hbar\, ( i \frac{d |b_k|}{dt} -  \, |b_k| \frac{d \phi_k }{dt})e^{i \phi_k }.
\label{derincmode}
\end{equation}
Next, replacing (\ref{derincmode}) in (\ref{ncmode}), multiplying this equation by $ e^{-i \phi_k }$ 
and separating the real and imaginary parts, one can, after some algebra, decouple Eq. (\ref{ncmode})
 into the following
 $ 2 N_c $ real equations, written in terms of  population $ n_k =|b_{k}|^2 = N_k / N $ and phase difference
$ \varphi_k= \phi_k - \phi_{k-1}$, 
\begin{eqnarray}
 \hbar\,\frac{dn_k}{dt}& = & - 2 J [ \sqrt{n_k \, n_{k+1}} \, \sin\varphi_{k+1} 
-\sqrt{n_k \, n_{k-1} } \, \sin\varphi_k ]\nonumber\\
&-&  J' N  [ \sqrt{n_k \, n_{k+1} } (n_k + n_{k+1} ) \, \sin\varphi_{k+1}
-\sqrt{n_k \, n_{k-1} } (n_k + n_{k-1} ) \, \sin\varphi_k]
\label{ncmode1hn}
\end{eqnarray}
\begin{eqnarray}
 \hbar\,\frac{d\varphi_k}{dt} & = & U N (n_{k-1} - n_k) \nonumber\\
&-&  J \left[ \left(\sqrt{\frac{n_k}{ n_{k-1}}} - \sqrt{\frac{n_{k-1} }{ n_k}}\,\right) \, \cos\varphi_k
+ \sqrt{\frac{n_{k-2}}{ n_{k-1} }} \, \cos\varphi_{k-1} 
- \sqrt{\frac{n_{k+1} }{ n_k}} \, \cos\varphi_{k+1}\right]\nonumber\\
&-& \frac{J'N}{2}\left[ \left( n_k \sqrt{\frac{n_k}{ n_{k-1} }} - n_{k-1} \sqrt{\frac{n_{k-1}}{ n_k}}\,\right)
 \, \cos\varphi_k 
+ \left(3\, \sqrt{n_{k-2} \, n_{k-1}} + n_{k-2} \sqrt{\frac{n_{k-2}}{ n_{k-1}}}\,\right) 
 \, \cos\varphi_{k-1} \right.\nonumber\\
&-& \left.\left(3\, \sqrt{n_{k+1} \, n_k} + n_{k+1} \sqrt{\frac{n_{k+1}} { n_k}}\,\right)  \, \cos\varphi_{k+1}\right].
\label{ncmode2hn}
\end{eqnarray}
The above MM dynamical equations 
constitute the generalization of the TM pair of equations (\ref{imb}) and (\ref{phase}) for
$N_c> 2$. Note that similarly to the TM case, only $2N_c- 2$ 
of the above equations  are  independent since the variables
must fulfill $\sum_k n_k=1$ and  $\sum_k \varphi_k=0$.
%

\subsection{Four-well ring lattice}\label{nc4}
In order to compare the MM dynamics with the results of GP simulations,
we have numerically integrated the system (\ref{ncmode1hn})-(\ref{ncmode2hn}) for $N_c=4$ and two initial configurations.
 The model parameters utilized in this case were
$ U = 1.38 \times 10^{-3} \, \hbar \omega_r $,
$ J =  -4.98\times 10^{-4} \, \hbar \omega_r $ and
$ J' =  2.76 \times 10^{-8} \, \hbar \omega_r $.

\subsubsection{Symmetric case}\label{nc4sc}

\begin{figure}
\includegraphics{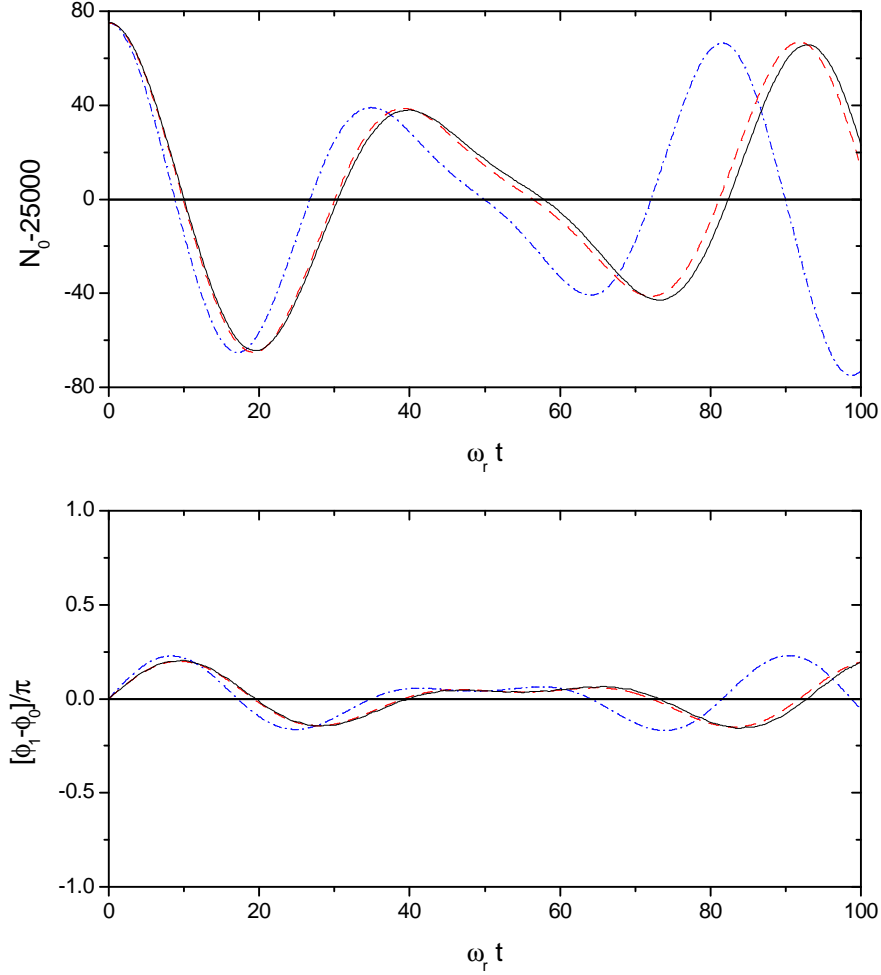}
\caption{(Color online) Time evolution of population $N_0 - M $ (top panel) and phase difference $ \phi_1-\phi_0 $ (bottom panel),
for the four-well system and an initial condition
$N_0 - M =75$, $N_k - M =-25$ ($k=1,2, 3$) with an uniformly vanishing phase. 
 The  solid line
corresponds to the GP simulation, while the dot-dashed (blue) and dashed (red) lines 
correspond to the MM and RMM models, with on-site interaction energy parameters  
$ U = 1.38 \times 10^{-3} \, \hbar \omega_r $ and
  $ U_R = 1.08 \times 10^{-3} \, \hbar \omega_r $, respectively.
}
\label{fig8}
\end{figure}
Here we consider initial conditions which are  symmetric with respect to the right and left 
from the $k=0$ well, as also studied by De Liberato and Foot in Ref. \cite{cuatropozos06}.
Particularly, in Fig. \ref{fig8}   we have chosen the following initial condition: 
$N_0 - M =75$ and $N_k - M =-25$ for the remaining sites, 
where $ M=N/N_c=25000 $
denotes the mean number of particles per site in the ground state.
We may observe in the top panel that the population
oscillates around the mean value $ M $
without any periodicity, at least for the times involved in our numerical simulations.
A similar behavior for the phase difference has been depicted in the bottom panel of Fig. \ref{fig8}.
Note that the MM model again reproduces the shape of the GP evolution in a
faster dynamics, as already observed for the TM model.
\begin{figure}
\includegraphics{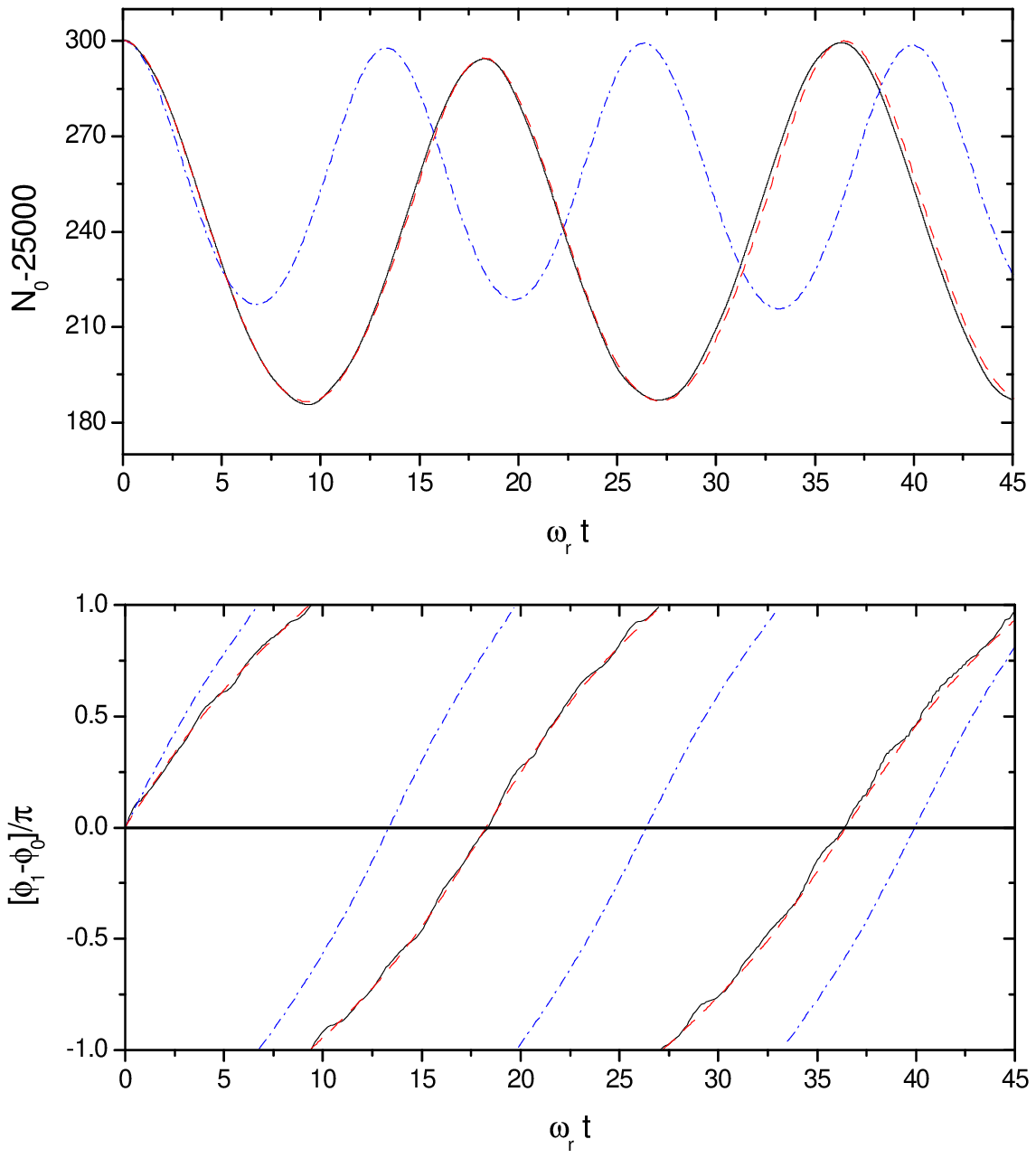}
\caption{(Color online) Same as Fig. \ref{fig8} for an initial condition
$N_0 - M =300$ and $N_k - M = -100$ ($k=1,2, 3$). }
\label{fig9}
\end{figure}
Figure \ref{fig9} shows the time evolution for the same symmetric initial configuration,
but with a
higher population in the $k=0$ well ($N_0 - M =300$, 
$N_k - M = - 100$ for $k=1,2,3$).
We observe in this case a clear  ST regime, with the population $N_0 - M$, which keeps positive during the
 oscillations performed around $N_0-25000\simeq 240$,
and an unbounded phase that increases almost linearly with time.

Now we will generalize the treatment of Sec. \ref{2-modeltimes}, to
estimate the 
ST period in order to derive a renormalized on-site energy parameter. 
First, according to the bottom panel of Fig. \ref{fig9} we may approximate (cf Eq. (\ref{fist}))
\begin{equation}
\varphi_1(t) \simeq \frac{2 \pi}{T_{\rm ST}} t,
\label{pphiaprox}
\end{equation}
and next, consistent with this approximation (cf Eq. (\ref{Iphaseap})), we approximate Eq. (\ref{ncmode2hn}) as,
\begin{equation}
 \frac{d\varphi_1}{dt} \simeq \frac{U N (n_{0} - n_1)}{\hbar}\simeq \frac{U (\overline{N_{0} - N_1 })}{\hbar},
\label{aproxi}
\end{equation}
where the upper bar again denotes time average. 
Therefore, from (\ref{pphiaprox}) and (\ref{aproxi}) we may estimate the ST period as, 
\begin{equation}
 T_{\rm ST}  \simeq  \frac { 2 \pi \hbar } { U (\overline{N_{0} - N_1 }) } 
\label{periodoaprox}
\end{equation}
which, taking into account the value $ ( \overline{N_{0} - N_1}) = 345 $ 
extracted from the MM results, yields 
$ T_{\rm ST}\simeq 13 \,\, \omega_r^{-1}$,
in accordance with the period of the MM model
in Fig. \ref{fig9} (dot-dashed (blue) lines).
Then, we may repeat the procedure of Sec. \ref{2-modeltimes} and extract a 
renormalized on-site interaction energy parameter,
\begin{equation}
 U_R   =  \frac { 2 \pi \hbar } { T_{\rm GP} (\overline{N_{0} - N_1})_{\rm GP}} \, ,
\label{urenormal}
\end{equation}
where the values $ (\overline{N_{0} - N_1})_{\rm GP}= 323 $ and 
$ T_{\rm GP} \simeq 18 \,\, \omega_r^{-1} $, arising from the GP simulation results, yield
$ U_R = 1.08 \times 10^{-3} \, \hbar \omega_r $.
The use of  this renormalized $U$ parameter in the MM calculations leads to
a much better agreement with the GP results, as clearly shown
in Figs. \ref{fig8} and \ref{fig9}. We will call this improved MM model as the 
renormalized multiple-mode (RMM) model.

\subsubsection{Non symmetric case }\label{nsc}

\begin{figure}
\includegraphics{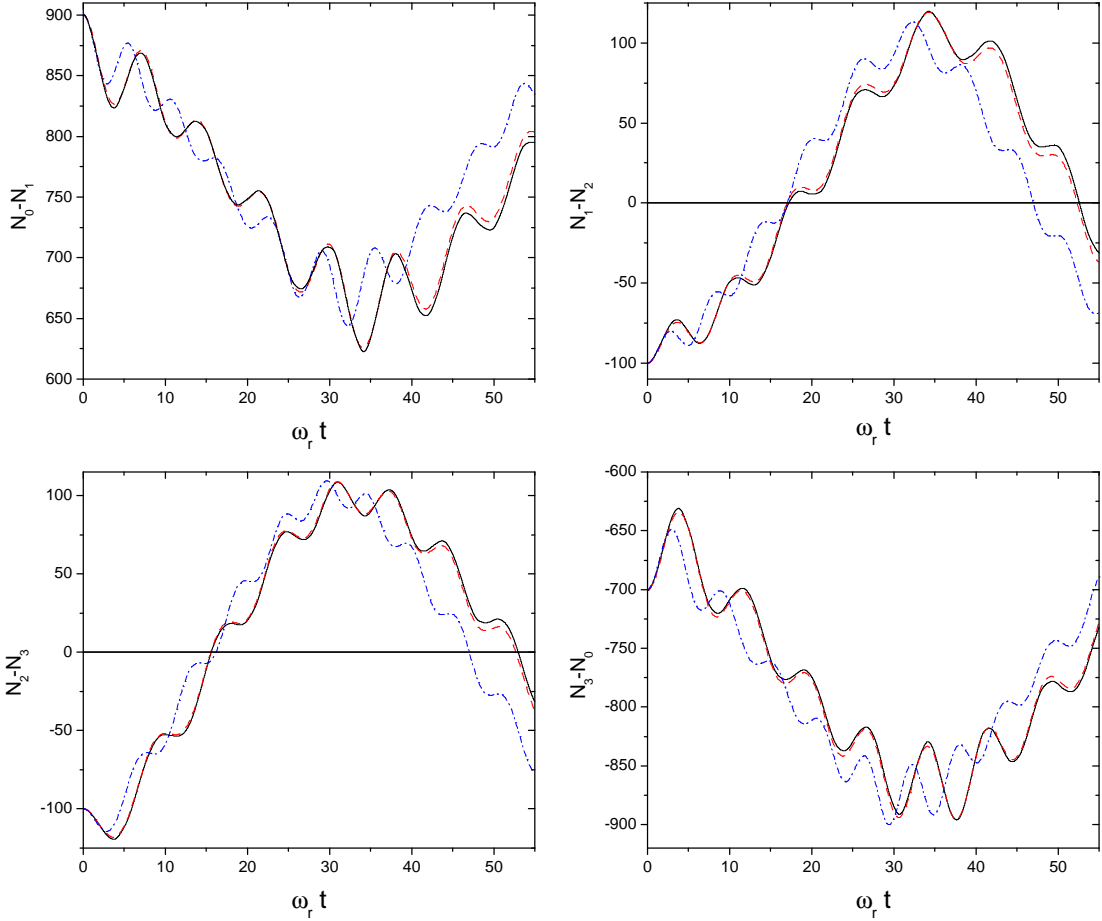}
\caption{ (Color online)  Particle number differences between neighboring sites for $N_c=4$
and the initial condition $N_0 - M =600$, $N_1 - M = -300$,
$ N_2 - M = -200 $, and $ N_3 - M = -100$.  The  solid line
corresponds to the GP simulation, while the dot-dashed (blue) and dashed (red) lines 
correspond to the MM and RMM models, with on-site interaction energy parameters  
$ U = 1.38 \times 10^{-3} \, \hbar \omega_r $ and
  $ U_R = 1.08 \times 10^{-3} \, \hbar \omega_r $, respectively.
 }
\label{asimd}
\end{figure}

\begin{figure}
\includegraphics{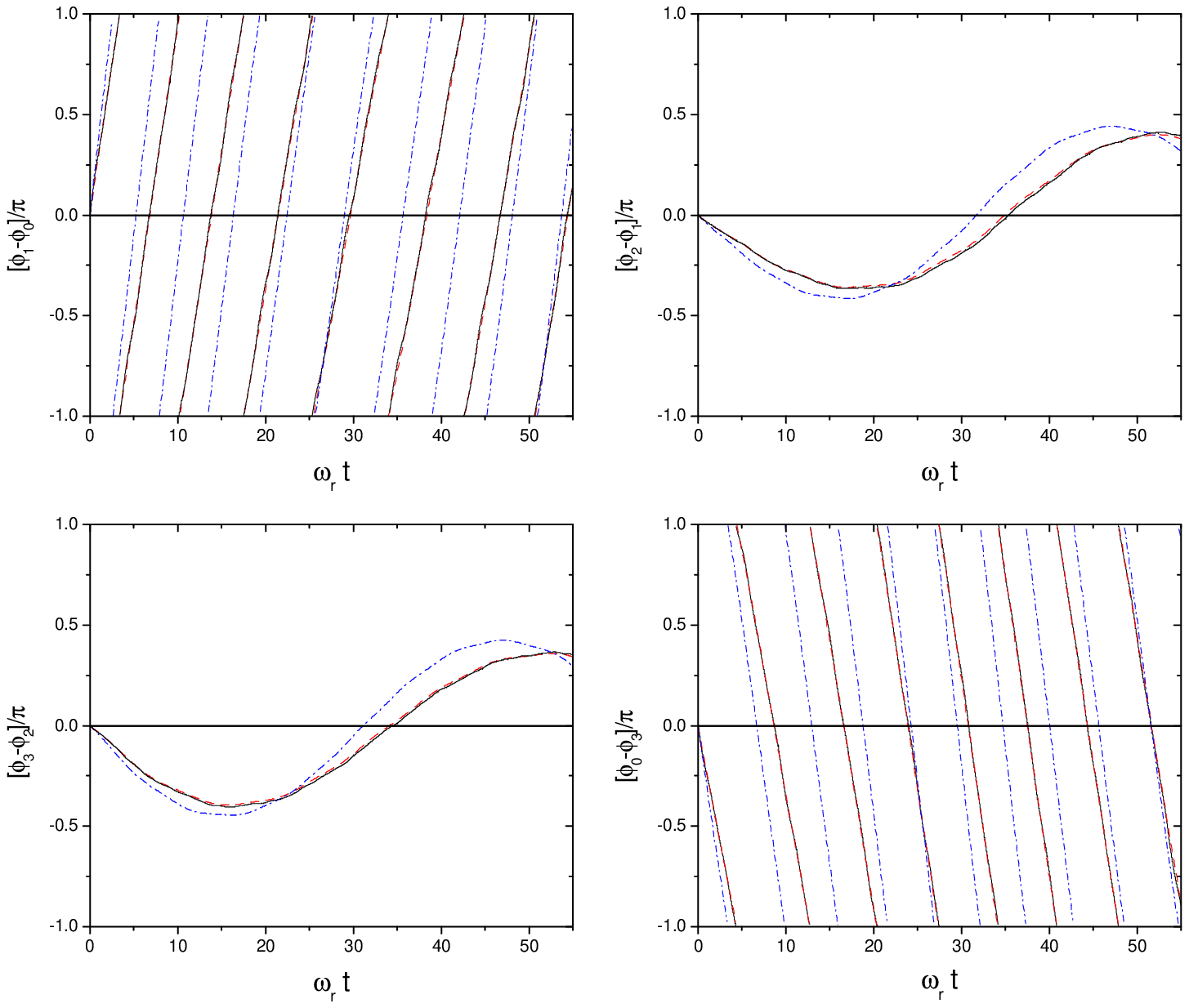}
\caption{ (Color online)  Same as Fig. \ref{asimd} for the  phase differences between neighboring sites.
 }
\label{asimf}
\end{figure}

\begin{figure}
\includegraphics{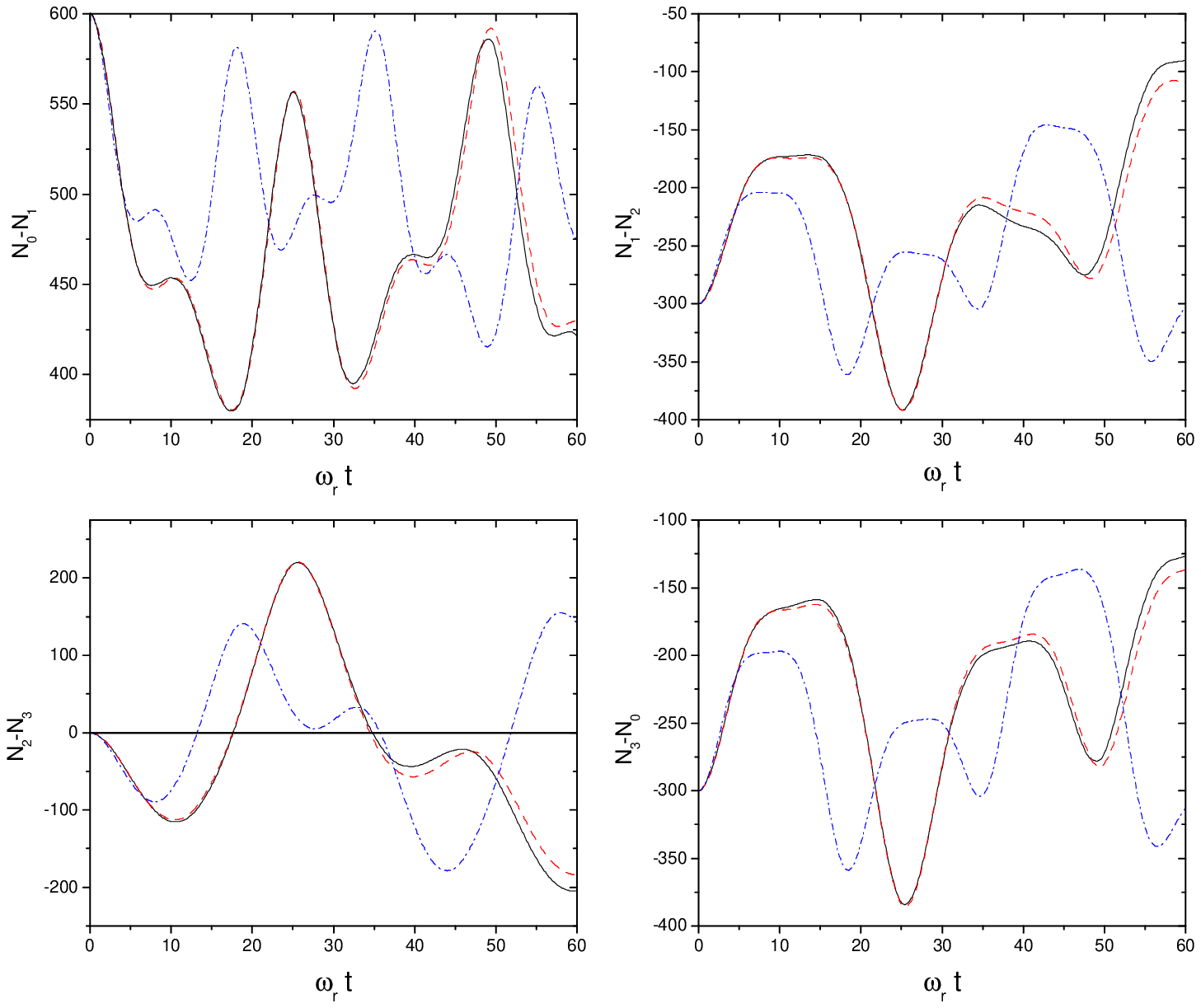}
\caption{ (Color online)  Same as Fig. \ref{asimd} for an initial condition $N_0 - M =300$, $N_1 - M = -300$,
$ N_k - M = 0 $ ($ k=2,3$).
 }
\label{fig7}
\end{figure}

\begin{figure}
\includegraphics{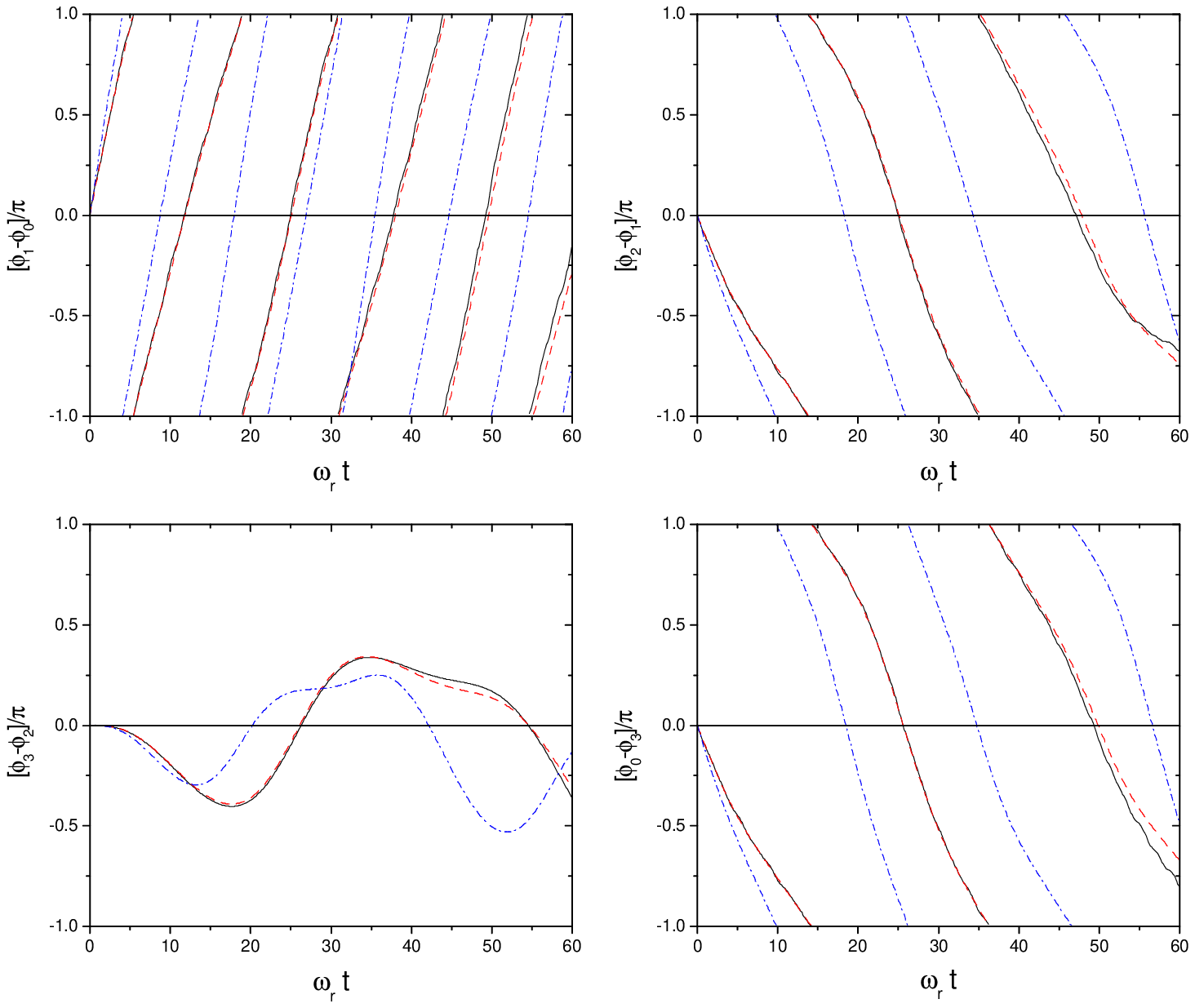}
\caption{ (Color online)  Same as Fig. \ref{fig7} for the  phase differences between neighboring sites.
 }
\label{fig8p}
\end{figure}

To test the quality of the above RMM model, we will analyze the time evolution of two
non symmetric initial configurations utilizing
the same value for $U_R$ extracted in the previous section.
In Figs. \ref{asimd} and \ref{asimf}, we have plotted the population and phase differences
between adjacent sites, respectively, for an initial condition $N_0 - M =600$, $N_1 - M = -300$,
$ N_2 - M = -200 $, and $ N_3 - M = -100$.
We may
observe that the RMM model fits much more accurately the GP simulation results  than the original MM model.
A similar improvement may be observed in Figs. \ref{fig7} and \ref{fig8p} for the second initial condition,
$N_0 - M =300$, $N_1 - M = -300$,
and $ N_k - M = 0 $ for $ k=2,3 $.
As inferred from Figs. \ref{asimd} and \ref{asimf}, such a configuration presents 
self-trapping in the $k=0$ site, while for the second initial condition,
this system exhibits  self-trapping in the $k=0$ site, self-depletion
in the $k=1$ site,
and an irregular oscillatory dynamics around the mean number of particles on the remaining wells,
as seen from Figs. \ref{fig7} and \ref{fig8p}.
The latter configuration had been previously described by means of a standard MM model by
De Liberato and Foot \cite{cuatropozos06}. 

\subsection{Eight-well ring lattice }\label{n8}

\begin{figure}
\includegraphics{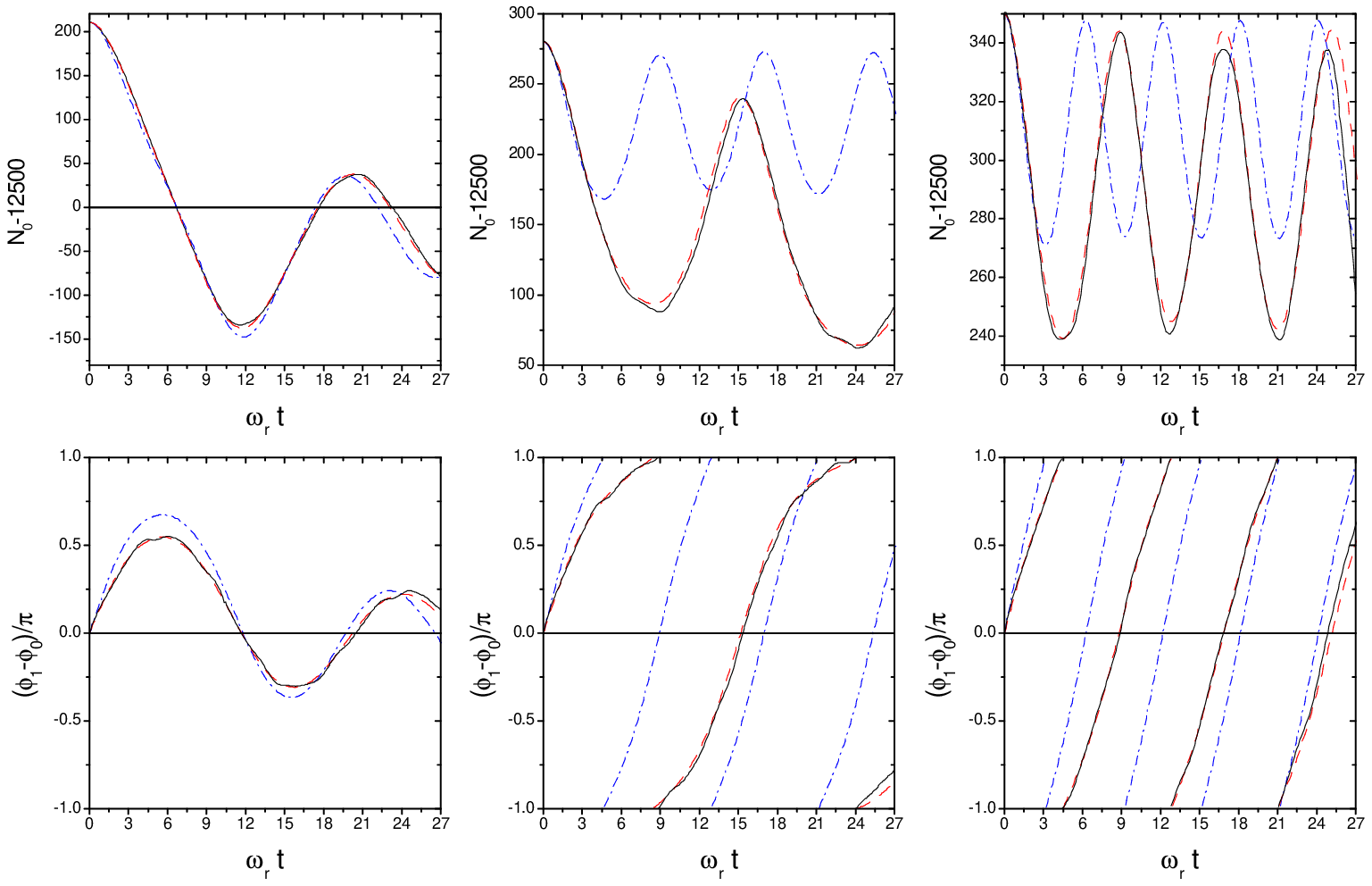}
\caption{ (Color online) Population of the $k=0$ well (upper panels)
 and phase difference 
between the $k=1$ and $k=0$ wells (lower panels)
for $N_c=8$ and the initial conditions: $N_0 - M =210$, $ N_k - M = -30 $ ($1\le k\le 7 $) (left panels); 
$N_0 - M =280$, $ N_k - M = -40 $ ($1\le k\le 7 $) (central panels) and
$N_0 - M =350$, $ N_k - M = -50 $ ($1\le k\le 7 $) (right panels).
The  solid line
corresponds to the GP simulation, while the dot-dashed (blue) and dashed (red) lines 
correspond to the MM and RMM models, with on-site interaction energy parameters  
$ U = 2.918 \times 10^{-3} \, \hbar \omega_r $ and
  $ U_R = 2.2 \times 10^{-3} \, \hbar \omega_r $, respectively.
}
\label{8porcion}
\end{figure}

To conclude we will explore a ring lattice consisting of a larger number of wells,
$N_c=8$.
The corresponding MM parameters are as follows,
$ U = 2.918 \times 10^{-3} \, \hbar \omega_r $, 
$ J = -1.898 \times 10^{-3} \, \hbar \omega_r $,
and $ J' = 2.118 \times 10^{-7} \, \hbar \omega_r $.
In  Fig. \ref{8porcion}, we have depicted the population of the site $k=0$ 
 and the  phase difference 
between the $k=1$ and $k=0$ sites,
for three different initial conditions.
Then, we may obtain as before a renormalized on-site energy parameter
$U_R$ from the GP results of the ST regime depicted on the
right panels of Fig. \ref{8porcion}.  In fact, replacing the ST period $T_{\rm GP}  \simeq 8.4 \,\, \omega_r^{-1}$
and the average difference $ (\overline{N_{0} - N_1})_{\rm GP} \simeq 335$ in Eq. (\ref{urenormal}),
%
%
we obtain a RMM model parameter $ U_R=2.2  \times 10^{-3}$, which yields
a sizable improvement to the MM results, as seen in  Fig. \ref{8porcion}.

\section{ Calculation of  the effective interaction energy parameter using the
 ground-state density}

Recently it has been demonstrated \cite{jez13,*cond} that for a double-well system,
the on-site interaction  energy dependence on the imbalance  should be taken into account  in  the two-mode model,
in order to accurately  describe the exact dynamics. 
There, using a Thomas-Fermi  density,
a linear dependence with the imbalance has been analytically encountered, and this
has been shown to
give rise to an effective  interaction energy parameter in 
the two-mode equations of motion. Here we generalize, beyond the Thomas-Fermi
approximation, that
result to the case of  multiple-well configurations.
Following the procedure of Ref. \cite{jez13,*cond} adapted to $N_c$ wells
and using numerically obtained densities,
we have to evaluate the quotient
\begin{DIFnomarkup}
\begin{equation}
\frac{U_k}{U}\simeq  \frac{ \int d^2{\bf r}\,\,  \rho_N({\bf r}) \,  \rho_{N+\Delta N}({\bf r})}
{ \int d^2{\bf r}\,\,  \rho^2_N({\bf r}) },
\label{coc}
\end{equation}
\end{DIFnomarkup}
where we have further assumed in (\ref{coc}) that instead of localized on-site densities, 
we may use the
ground-state densities $\rho_N({\bf r})$ and $\rho_{N+\Delta N}({\bf r})$ normalized to unity
of systems with $N$ and $N + \Delta N $ particles, respectively,  being 
$ \Delta N=  N_c  \Delta N_k= N_c N_k-N$. 

\begin{figure}
\includegraphics{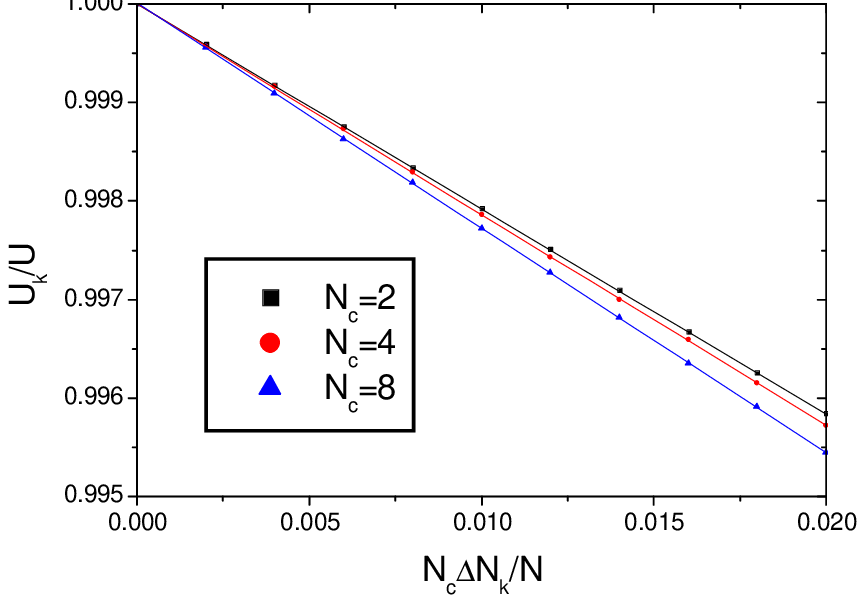}
\caption{ (Color online) Numerical calculation of the
right-hand side of Eq. (\ref{coc}) versus $ N_c \Delta N_k / N $
for three numbers of lattice sites. The (black) squares, (red) circles, and (blue) triangles correspond
to $N_c=2$, 4, and 8, respectively, while each line corresponds to a linear fit.
}
\label{figult}
\end{figure}

Numerical calculations of the r.h.s. of (\ref{coc}) 
 are depicted in Fig. \ref{figult}, 
where we may observe a linear behavior with
$ N_c \Delta N_k / N $
for different numbers 
of lattice sites.
Note that the apparent counterintuitive decrease of  this function  with the site population is related  to the fact that the densities
must be  normalized to unity.

Taking into account this  linear dependence of the on-site energy parameter we may write
\begin{DIFnomarkup}
\begin{equation}
\frac{U_{k}}{U}  \simeq  1  -   \alpha   \frac{ N_c  \Delta N_k} {N},
\label{Uek} 
\end{equation}
where the values of $\alpha $ in Table \ref{table} correspond to the linear fits of the points in Fig. \ref{figult}.
To include this correction in the MM model 
 we must evaluate \cite{jez13,*cond},
\begin{equation}
\frac{U_{k-1}}{U} N_{k-1}  -  \frac{ U_{k} }{U}  N_{k}  = \left(1-\alpha  \frac{ N_c  \Delta N_{k-1}} {N}\right)\left(  \Delta N_{k-1} + \frac{ N}{N_c}\right)
- \left(1-\alpha  \frac{ N_c  \Delta N_{k}} {N}\right) \left( \Delta N_{k} +  \frac{N}{N_c}\right)    
\label{resta}
\end{equation}
which yields,
\begin{equation}
\frac{U_{k-1}}{U} N_{k-1}  -  \frac{ U_{k} }{U}  N_{k}  = (1-\alpha)  ( N_{k-1} -N_{k})   -  
 \alpha (  N_{k-1} - N_{k}) [  \frac{ N_c (N_{k-1}+  N_{k})}{ N}-2 ].   
\label{restaf}
\end{equation}
And finally replace the last result in the first term of the r.h.s. of Eqs. (\ref{phase}) and (\ref{ncmode2hn}). 
By analyzing the term
\begin{equation}
 \alpha (  N_{k-1} - N_{k}) [  \frac{ N_c (N_{k-1}+  N_{k})}{ N}-2 ] =
 \alpha (  N_{k-1} - N_{k})  \frac{ N_c (\Delta N_{k-1}+ \Delta N_{k})}{ N},
\label{error}
\end{equation}
\end{DIFnomarkup}
we first note that in the double well case it is identically zero, while for other
 studied cases
$ \frac{ N_c (\Delta N_{k-1}+ \Delta N_{k})}{ N} <<1 $ (cf. the range
of $ N_c \Delta N_{k}/ N$ in Fig. \ref{figult}). 
Thus, disregarding such a term, we in fact obtain a correction that can be regarded 
as a reduced 
 effective interaction parameter $ \tilde{U}=  (1-\alpha)  U $.
Note that this result is in accordance with our previous analysis which 
yielded the renormalized parameter $U_R$
using characteristic times, while the corresponding quantitative agreement is shown 
in Table \ref{table}. 

\begin{table}
\caption{ Linear correction coefficient $\alpha$ of the on-site interaction energy 
parameter of  the $k$-site  (Eq. (\ref{Uek})), and 
  effective $\tilde{U}$, renormalized $U_R$, and bare $U$ interaction energy parameters, 
for three numbers of wells $ N_c$. 
The interaction  energy  parameters are given  in units of $\hbar \omega_r$.
}
\begin{ruledtabular}
\begin{tabular}{lcccc}
$ N_c$ & $   \alpha $ &   $ \tilde{U}$   &         $ U_R$   &  $ U $  \\[3pt]
\hline \\[-5pt]
2  &  $  0.208 $ &   $  5.33 \times 10^{-4}   $ &  $   5.28 \times 10^{-4}   $ &  $   6.73 \times 10^{-4}   $  \\[3pt]
4 &  $  0.214 $ &    $    1.09  \times 10^{-3}  $ &   $    1.08  \times 10^{-3}  $ &  $   1.38  \times 10^{-3}   $ \\[3pt]
8  &  $  0.228 $ &    $ 2.25 \times 10^{-3} $ &  $   2.22 \times 10^{-3} $  &  $   2.92 \times 10^{-3}   $ \\[3pt]
\end{tabular}
\end{ruledtabular}
\label{table}
\end{table}

\section{ Summary and concluding remarks}\label{Conclusions} 

We have investigated the dynamics of ring-shaped optical lattices 
with a high number of particles per site.
To this aim, we have derived the equations of motion
for population 
and phase differences between neighboring sites  of a generalized multimode model
that utilizes a localized on-site Wannier-like
 basis.
We have shown that in case of a double-well system,
this approach coincides with the latest
improved two-mode model \cite{anan06}.

To test the quality of our model,
we have numerically solved the time dependent GP equation for different numbers of wells, 
particularly $2$, $4$, and $8$. 
By realizing  that the self-trapping time  period turns out to be chiefly
ruled by the on-site interaction
energy parameter, and utilizing the output of a single GP simulation,
we were able to renormalize such a parameter. 
The use of this renormalized parameter in the multimode equations strikingly led 
to a much better agreement with the GP results for all investigated initial conditions,
of which only a few representative were included in this report.
Finally, we have shown that the 
effective interaction energy parameter,
which takes into account the deformation of the density during the time evolution, 
yields results 
that are in good agreement with the previously obtained for the renormalized parameter.

To conclude, we wish to emphasize that the two-mode model has predicted,
even in its improved version \cite{anan06},
a sizable faster evolution than
our GP simulation results, as discussed in Sec. \ref{2-modeltimes}.
The same behavior is observed in previous experimental
and theoretical works dealing with other type of double-well systems
(see, e.g., \cite{mele11,gati} and references therein). 
We believe that also in these systems as in our case, the TM model with an effective
reduction of the on-site interaction energy parameter numerically
calculated as here proposed, should provide
a more accurate dynamics.


\acknowledgments
We acknowledge M. Guilleumas for a careful reading of the manuscript.
DMJ and HMC acknowledge CONICET for financial support under Grants PIP Nos. 
11420090100243 and 11420100100083, respectively.

\providecommand{\noopsort}[1]{}\providecommand{\singleletter}[1]{#1}%

\end{document}